\documentclass[preprint,journal]{vgtc} 
\ifpdf
  \pdfoutput=1\relax
  \usepackage{graphicx}
  \DeclareGraphicsExtensions{.pdf,.png,.jpg,.jpeg}
  \usepackage{graphicx}
  \DeclareGraphicsExtensions{.eps}
\fi
\graphicspath{{figures/}{pictures/}{images/}{./}}
\usepackage{microtype}
\PassOptionsToPackage{warn}{textcomp}
\usepackage{textcomp}
\usepackage{amsmath}
\usepackage{amsfonts}
\usepackage{amsthm}
\usepackage[ruled,linesnumbered]{algorithm2e}
\usepackage{times}
\usepackage{xcolor}
\usepackage{enumitem}
\definecolor{classred}{HTML}{ff4757}
\definecolor{classblue}{HTML}{70a1ff}
\definecolor{poisonred}{HTML}{ffa502}
\definecolor{poisonblue}{HTML}{49C17A}
\usepackage{cite}                      
\usepackage{tabu}                      
\usepackage{booktabs}                  
\onlineid{1096}
\vgtccategory{Research}
\vgtcpapertype{application/design study}
\title{Explaining Vulnerabilities to Adversarial Machine Learning through Visual Analytics}
\author{Yuxin Ma, Tiankai Xie, Jundong Li, Ross Maciejewski, \textit{Senior Member, IEEE}}
\authorfooter{
\item
\vspace{-1.2mm}
Y. Ma, T. Xie and R. Maciejewski are with the School of Computing, Informatics \& Decision Systems Engineering, Arizona State University. E-mail: \{yuxinma,txie21,rmacieje\}@asu.edu.
\item
J. Li is with the Department of Electrical and Computer Engineering, University of Virginia. E-mail: jl6qk@virginia.edu.
}
\abstract{
Machine learning models are currently being deployed in a variety of real-world applications where model predictions are used to make decisions about healthcare, bank loans, and numerous other critical tasks. As the deployment of artificial intelligence technologies becomes ubiquitous, it is unsurprising that adversaries have begun developing methods to manipulate machine learning models to their advantage. While the visual analytics community has developed methods for opening the black box of machine learning models, little work has focused on helping the user understand their model vulnerabilities in the context of adversarial attacks. In this paper, we present a visual analytics framework for explaining and exploring model vulnerabilities to adversarial attacks. Our framework employs a multi-faceted visualization scheme designed to support the analysis of data poisoning attacks from the perspective of models, data instances, features, and local structures. We demonstrate our framework through two case studies on binary classifiers and illustrate model vulnerabilities with respect to varying attack strategies. 
}
\teaser{
  \centering
  \includegraphics[width=\linewidth]{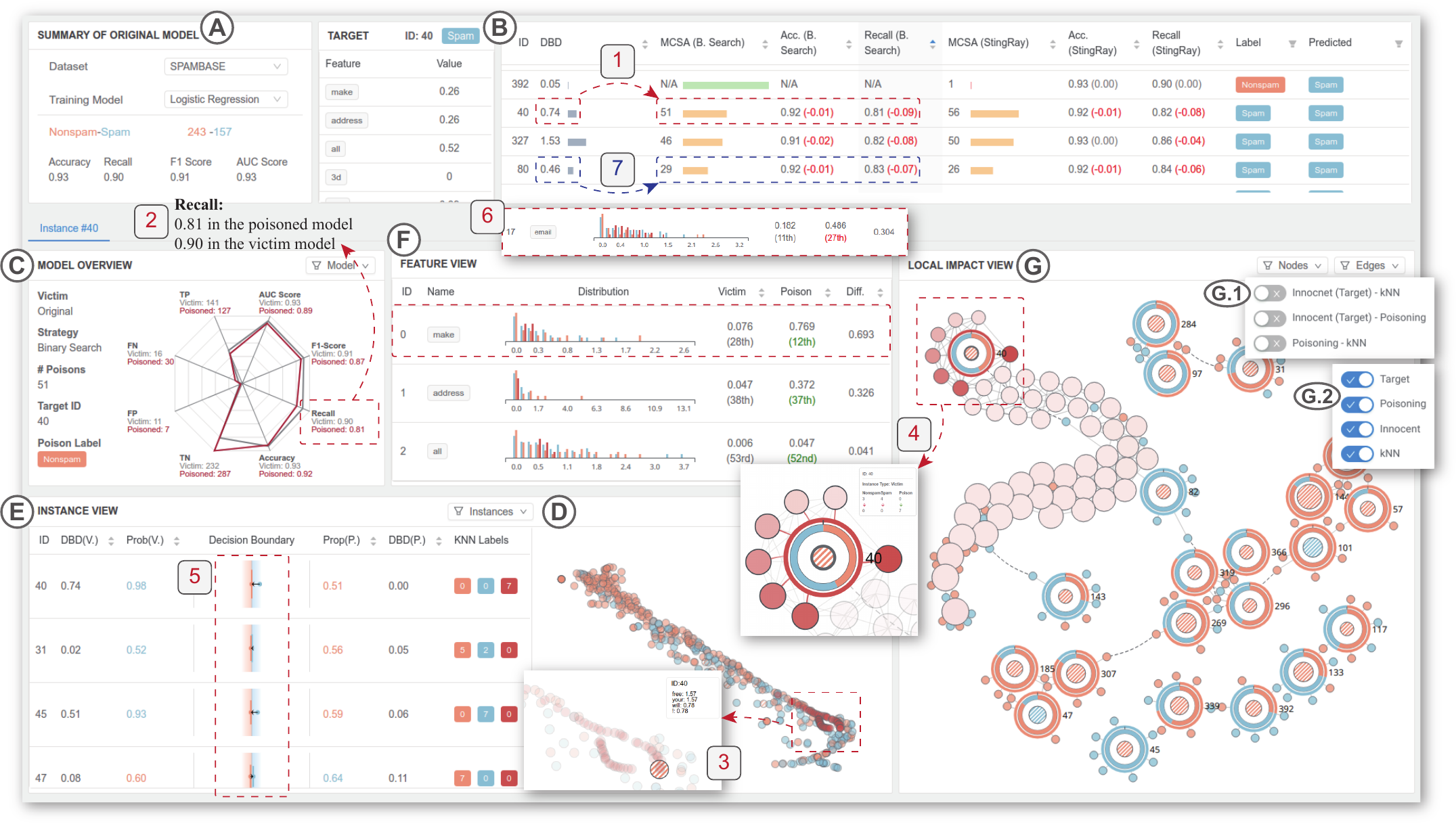}
  \caption{Reliability attack on spam filters. (1) Poisoning instance \#40 has the largest impact on the recall value, which is (2) also depicted in the model overview. (3) There is heavy overlap among instances in the two classes as well the poisoning instances. (4) Instance \#40 has been successfully attacked causing a number of innocent instances to have their labels flipped. (5) The flipped instances are very close to the decision boundary. (6) On the feature of words ``will'' and ``email'', the variances of poisoning instances are large. (7) A sub-optimal target (instance \#80) has less impact on the recall value, but the cost of insertions is 40\% lower than that of instance \#40.}
	\label{fig:teaser}
}
\vgtcinsertpkg
\ieeedoi{10.1109/TVCG.2019.2934631}

\begin{document}

\maketitle
\section{Introduction}
In the era of Big Data, Artificial Intelligence and Machine Learning have made immense strides in developing models and classifiers for real-world phenomena. To date, applications of these models are found in cancer diagnosis tools~\cite{esteva2017dermatologist}, self-driving cars~\cite{martinez2018driving}, biometrics~\cite{sundararajan2018deep}, and numerous other areas. Many of these models were developed under assumptions of static environments, where new data instances are assumed to be from a statistical distribution similar to that of the training and test data. Unfortunately, the real-world application of these models introduces a dynamic environment which is home to malicious individuals who may wish to exploit these underlying assumptions in the machine-learning models. Consider e-mail spam filtering as an example. To date, a variety of machine learning methods~\cite{blanzieri2008survey,caruana2012survey} have been developed to protect e-mail inboxes from unwanted messages. These methods build models to classify e-mail as spam or not-spam. However, adversaries still want their spam messages to reach your inbox, and these adversaries try to build input data (i.e., spam e-mails) that will fool the model into classifying their spam as safe. This can be done by misspelling words that might cause the machine learning classifier to flag a mail as spam or by inserting words and phrases that might cause the classifier to believe the message is safe.
Other adversarial attacks have explored methods to fake bio-metric data to gain access to personal accounts~\cite{biggio2015adversarial} and to cause computer vision algorithms to misclassify stop signs~\cite{chen2018shapeshifter}. Such exploits can have devastating effects, and researchers are finding that applications of machine learning in real-world environments are increasingly vulnerable to adversarial attacks. As such, it is imperative that model designers and end-users be able to diagnose security risks in their machine learning models. 

Recently, researchers have begun identifying design issues and research challenges for defending against adversarial machine learning, such as data de-noising, robust modeling, and defensive validation schemes ~\cite{Adv2018book,biggio2018wild}, citing the need to identify potential vulnerabilities and explore attack strategies to identify threats and impacts. These challenges lend themselves well to a visual analytics paradigm, where training datasets and models can be dynamically explored against the backdrop of adversarial attacks. In this paper, we present a visual analytics framework (Figure 1) designed to explain model vulnerabilities with respect to adversarial attack algorithms. Our framework uses modularized components to allow users to swap out various attack algorithms. A multi-faceted visualization scheme summarizes the attack results from the perspective of the machine learning model and its corresponding training dataset, and coordinated views are designed to help users quickly identify model vulnerabilities and explore potential attack vectors. For an in-depth analysis of specific data instances affected by the attack, a locality-based visualization is designed to reveal neighborhood structure changes due to an adversarial attack. 
To demonstrate our framework, we explore model vulnerabilities to data poisoning attacks. Our contributions include:
\begin{itemize}
    \item A visual analytics framework that supports the examination, creation, and exploration of adversarial machine learning attacks;
    \item A visual representation of model vulnerability that reveals the impact of adversarial attacks in terms of model performance, instance attributes, feature distributions, and local structures.
\end{itemize}

\section{Related Work}
Our work focuses on explaining model vulnerabilities in relation to adversarial attacks. In this section, we review recent work on explainable artificial intelligence and adversarial machine learning.

\subsection{Explainable Artificial Intelligence - XAI}
Due to the dramatic success of machine learning, artificial intelligence applications have been deployed into a variety of real-world systems. However, the uptake of these systems has been hampered by the inherent black-box nature of these machine learning models~\cite{krause2016interacting}. Users want to know why models perform a certain way, why models make specific decisions, and why models succeed or fail in specific instances~\cite{Endert2017}. The visual analytics community has tackled this problem by developing methods to open the black-box of machine learning models~\cite{Bertini2009,LIU201748,Lu2017,Lujunhua2017}. The goal is to improve the explainability of models, allow for more user feedback, and increase the user's trust in a model. To date, a variety of visual analytics methods have been developed to support model explainability and performance diagnosis.

\textbf{Model Explainability:} Under the black-box metaphor of machine learning, several model-independent approaches have been developed in the visual analytics community.  EnsembleMatrix~\cite{Talbot2009} supports the visual adjustment of preferences among a set of base classifiers. Since the base classifiers share the same output protocol (confusion matrices), the approach does not rely on knowledge of specific model types. 

In EnsembleMatrix, the users are provided a visual summary of the model outputs to help generate insights into the classification results. The RuleMatrix system~\cite{Ming2018} also focuses on the input-output behavior of a classifier through the use of classification rules, where a matrix based-visualization is used to explain classification criterion. Similarly, model input-output behaviors were utilized in Prospector~\cite{krause2016interacting}, where the relations between feature values and predictions are revealed by using partial dependence diagnostics. 

While those approaches focused on utilizing model inputs and outputs, other visual analytics work focuses on ``opening the black box,'' utilizing the internal mechanisms of specific models to help explain model outputs. Work by Muhlbacher et al.~\cite{Muhlbacher2014} summarizes a set of guidelines for integrating visualization into machine learning algorithms through a formalized description and comparison. For automated iterative algorithms, which are widely used in model optimization, Muhlbacher et al. recommended exposing APIs so that visualization developers can access the internal iterations for a tighter integration of the user in the decision loop. In terms of decision tree-based models, BaobabView~\cite{VandenElzen2011} proposes a natural visual representation of decision tree structures where decision criterion are visualized in the tree nodes. BOOSTVis\cite{Liu2017b} and iForest\cite{Zhao2019} also focus on explaining tree ensemble models through the use of multiple coordinated views to help explain and explore decision paths. Similarly, recent visual analytics work on deep learning~\cite{Yosinski2015,Kahng,Rauber2016,Ming2017,Wongsuphasawat2017,Kahng2017a,Liu2018,Wang2018,Wang2018a,Kwon2018,Strobelt2019} tackles the issue of the low interpretability of neural network structures and supports revealing the internal logic of the training and prediction processes. 

\textbf{Model Performance Diagnosis:} It is also critical for users to understand statistical performance metrics of models, such as accuracy and recall. These metrics are widely-used in the machine learning community to evaluate prediction results; however, these metrics provide only a single measure, obfuscating details about critical instances, failures, and model features~\cite{Amershi,Zhang2018}. To better explain performance diagnostics, a variety of visual analytics approaches have been developed. Alsallakh et al.\cite{Alsallakh2014} present a tool for diagnosing probabilistic classifiers through a novel visual design called Confusion Wheel, which is used as a replacement for traditional confusion matrices.

For multi-class classification models, Squares\cite{ren2017squares} establishes a connection between statistical performance metrics and instance-level analysis with a stacked view. Zhang et al.~\cite{Zhang2018} propose Manifold, a model-agnostic framework that does not rely on specific model types; instead, Manifold analyzes the input and output of a model through an iterative analysis process of inspection, explanation, and refinement. Manifold supports a fine-grained analysis of ``symptom'' instances where predictions are not agreed upon by different models. Other work has focused on profiling and debugging deep neural networks, such as LSTMs~\cite{Strobelt2016a}, sequence-to-sequence models~\cite{Strobelt2019}, and data-flow graphs~\cite{Wongsuphasawat2017}.

While these works focus on linking performance metrics to input-output instances, other methods have been developed for feature-level analysis to enable users to explore the relations between features and model outputs. 
For example, the INFUSE system~\cite{Krause2014} supports the interactive ranking of features based on feature selection algorithms and cross-validation performances. Later work by Krause et al.~\cite{Krause2017} also proposed a performance diagnosis workflow where the instance-level diagnosis leverages measures of ``local feature relevance'' to guide the visual inspection of root causes that trigger misclassification. 

As such, the visual analytics community has focused on explainability with respect to model input-outputs, hidden layers, underlying ``black-box'' mechanisms, and performance metrics; however, there is still a need to explain model vulnerabilities. To this end, Liu et al.~\cite{Liu2018VAST} present AEVis, a visual analytics tool for deep learning models, which visualizes data-paths along the hidden layers in order to interpret the prediction process of adversarial examples. However, the approach is tightly coupled with generating adversarial examples for deep neural networks, which is not extensible to other attack forms and model types. Our work builds upon previous visual analytics explainability work, adopting coordinated multiple views that support various types of models and attack strategies. What is unique in our work is the integration of attack strategies into the visual analytics pipeline, which allows us to highlight model vulnerabilities.

\subsection{Adversarial Machine Learning}
\label{sub:related_work_adv_learning}

\begin{figure}[ht]
	\centering	
	\includegraphics[width=1.00\columnwidth]{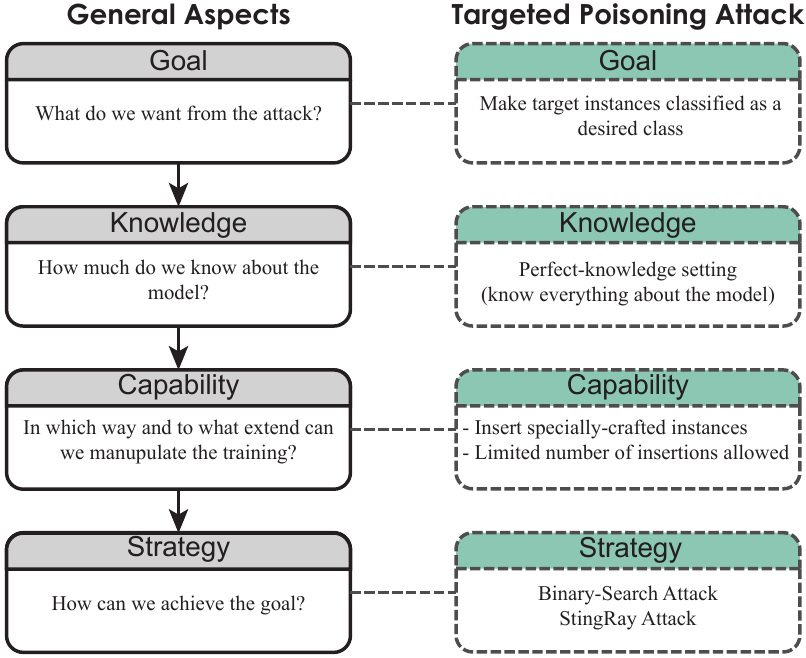}
	\caption{Key features of an adversary. (Left) The general components an adversary must consider when planning an attack. (Right) Specific considerations in a data poisoning attack.}
	\label{fig:data_poisoning_pipeline}
	\vspace{-4mm}
\end{figure}

Since our goal is to support the exploration of model vulnerabilities, it is critical to identify common attack strategies and model weaknesses.
The four main features of an adversary (or attacker)~\cite{Adv2018book,biggio2018wild} are the adversary's Goal, Knowledge, Capability, and Strategy, Figure~\ref{fig:data_poisoning_pipeline} (Left).

\vspace{1mm} \noindent \textbf{Goal:} 
In adversarial machine learning, an attacker's goal can be separated into two major categories: \textit{targeted attacks} and \textit{reliability attacks}. In a targeted attack, the attacker seeks to insert specific malicious instances or regions in the input feature space and prevent these insertions from being detected~\cite{Trojannn,shafahi2018poison}. In a reliability attack, the goal of the attacker is to maximize the overall prediction error of the model and make the model unusable for making predictions~\cite{steinhardt2017certified}. 

\vspace{1mm} \noindent \textbf{Knowledge:} The information that can be accessed by an attacker plays a significant role in how an attacker will design and deploy attack operations. The more knowledge an attacker has about a model (victim), the more precise an attack can be. In a \textit{black-box} model, the attacker will have imprecise (or even no) knowledge about the machine learning model, while in a \textit{white-box} setting, the attacker will have most (if not all) of the information about the model, including the model type, hyper-parameters, input features, and training dataset~\cite{biggio2018wild}. 

\vspace{1mm} \noindent \textbf{Capability:} The capability of the attacker refers to when and what the attacker can do to influence the training and testing process to achieve the attack's goal. Where the attack takes place (i.e., the stage of the modeling process - training, testing) limits the capability of the attacker. For example, \textit{poisoning attacks}~\cite{Xiao:2012:ALF:3007337.3007488,Biggio:2012:PAA:3042573.3042761} take place during the training-stage, and the attacker attempts to manipulate the training dataset. Typical operations in data poisoning attacks include adding noise instances and flipping labels of existing instances. An \textit{evasion attack}~\cite{dalvi2004adversarial,10.1007/978-3-642-40994-3_25,43405} takes place during the testing stage. Such an attack is intended to manipulate unlabeled data in order to avoid detection in the testing stage without touching the training process. In all of these cases, the attacker is constrained by how much they can manipulate either the training or test data without being detected or whether the training and test data are even vulnerable to such attacks.

\vspace{1mm} \noindent \textbf{Strategy:} Given the attacker's goal, knowledge, and capabilities, all that remains is for the attacker to design an attack strategy. An optimal attack strategy can be described as maximizing the attack effectiveness while minimizing the cost of data manipulation or other constraints~\cite{mei2015using}. 

\vspace{1mm} Currently, numerous adversarial machine learning attacks are being developed, with evasion and poisoning strategies receiving the most attention~\cite{AML-Review}.
In evasion attacks, a common strategy is to add noise to test data instances. Goodfellow et al.~\cite{43405} proposed a method to add ``imperceptible'' noise to an image, which can drastically confuse a trained deep neural network resulting in unwanted predictions. For poisoning attacks, the strategies are usually formalized as bi-level optimization problems, such as gradient ascending~\cite{Biggio:2012:PAA:3042573.3042761} and machine teaching~\cite{mei2015using}.
Common among these attacks is the goal of manipulating the trained model, and it is critical for users to understand where and how their models may be vulnerable.

\section{Design Overview}

Given the key features of an adversary, we have designed a visual analytics framework that uses existing adversarial attack algorithms as mechanisms for exploring and explaining model vulnerabilities.
Our framework is designed to be robust to general adversarial machine learning attacks. However, in order to demonstrate our proposed visual analytics framework, we focus our discussion on \textbf{targeted data poisoning attacks}~\cite{biggio2018wild}. Data poisoning is an adversarial attack that tries to manipulate the training dataset in order to control the prediction behavior of a trained model such that the model will label malicious examples into a desired classes (e.g., labeling spam e-mails as safe). Figure~\ref{fig:data_poisoning_pipeline} (Right) maps the specific goal, knowledge, capabilities, and strategies of a poisoning attack to the generalized adversarial attack.   

For the purposes of demonstrating our framework, we assume that the attack takes place in a white-box setting, i.e., the attacker has full knowledge of the training process. Although the scenario seems partial to attackers, it is not unusual for attackers to gain perfect- or near-perfect-knowledge of a model by adopting multi-channel attacks through reverse engineering or intrusion attacks on the model training servers~\cite{biggio2014security}. Furthermore, in the paradigm of proactive defense, it is meaningful to use the worst case attack to explore the upper bounds of model vulnerability~\cite{biggio2018wild}. In terms of poisoning operations on the training dataset, we focus on causative attacks~\cite{Barreno2010}, where attackers are only allowed to insert specially-crafted data instances. This kind of insertion widely exists in real-world systems, which need to periodically collect new training data, examples include recommender systems and email spam filters~\cite{steinhardt2017certified}. In such attacks, there is a limit to the number of poisoned instances that can be inserted in each attack iteration, i.e., a budget for an attack. An optimal attack attempts to reach its goal by using the smallest number of insertions within the given budget.

\subsection{Analytical Tasks}
After reviewing various literature on poisoning attacks~\cite{Biggio:2012:PAA:3042573.3042761,Mozaffari2015Systematic,steinhardt2017certified,shafahi2018poison,suciu2018does,jagielski2018manipulating,Adv2018book,AML-Review,biggio2018wild}, we extracted common high-level tasks for analyzing poisoning attack strategies. These tasks were refined through discussions with our co-author, a domain-expert in adversarial machine learning.

\vspace{2mm} \noindent \textbf{T1 Summarize the attack space.} A prerequisite for many of the algorithms is to set target instances to be attacked in the training dataset. In our framework, analysts need to be able to identify attack vectors and vulnerabilities of the victim model in order to specify target instances.

\vspace{2mm} \noindent \textbf{T2 Summarize the attack results.} By following the well-known visual information seeking mantra~\cite{Ben1996mantra}, the system should provide a summary of the attack results after an attack is executed. In data poisoning, typical questions that the attackers might ask include:
\begin{itemize}
    \item \textbf{T2.1} How many poisoning data instances are inserted? What is their distribution? Has the attack goal been achieved yet?
    \item \textbf{T2.2} What is the performance of the model before and after the attack and is there a significant difference? How many instances in the training dataset are misclassified by the poisoned model?
\end{itemize}
 
\noindent \textbf{T3 Diagnose the impact of data poisoning.} In this phase, the user explores the prediction results and analyzes the details of the poisoned model. Inspired by the recent work in interpretable machine learning~\cite{Alsallakh2014,ren2017squares,Krause2017,Zhang2018}, we explore the influence of insertion focusing on: attribute changes for individual instances; and drifts of data distributions on features due to poisoning. We consider both instance-level and feature-level diagnoses when investigating the impact of poisoning data. The following questions are explored in this phase:

\begin{itemize}
    \item \textbf{T3.1} At the instance-level, is the original prediction different from the victim model prediction? How close is the data instance to the decision boundary? How do the neighboring instances affect the class label? Is there any poisoned data in the data-instance's top-k nearest neighbors? 
    \item \textbf{T3.2} At the feature-level, what is the impact of data poisoning on the feature distributions? 
\end{itemize}

\subsection{Design Requirements}
From the task requirements, we iteratively refined a set of framework design requirements to identify how visual analytics can be used to best to support attack analysis and explanation. We have mapped different analytic tasks to each design requirement. 

\vspace{2mm} \noindent \textbf{Visualizing the Attack Space - D1.} The framework should allow users to upload their victim model and explore vulnerabilities. By examining statistical measures of attack costs and potential impact, the users should be able to find weak points in the victim model depending on the application scenario, and finally identify desired target instances for in-depth analysis in the next step (T1).

\vspace{2mm} \noindent \textbf{Visualizing Attack Results - D2.} To analyze the results of an attack, the framework should support overview and details-on-demand:
\begin{itemize}
    \item \textit{Model Overview - D2.1}, summarize prediction performance for the victim model as well as the poisoned model (T2.2);
    \item \textit{Data Instances - D2.2}, present the labels of the original and poisoned data instances (T2.1, T3.1);
    \item \textit{Data Features - D2.3}, visualize the statistical distributions of data along each feature (T3.2);
    \item \textit{Local Impacts - D2.4}, depict the relationships between target data instances and their nearest neighbors (T3.1).
\end{itemize}

\section{Visual Analytics Framework}
\begin{figure}[ht]
	\centering	
	\includegraphics[width=1.00\columnwidth]{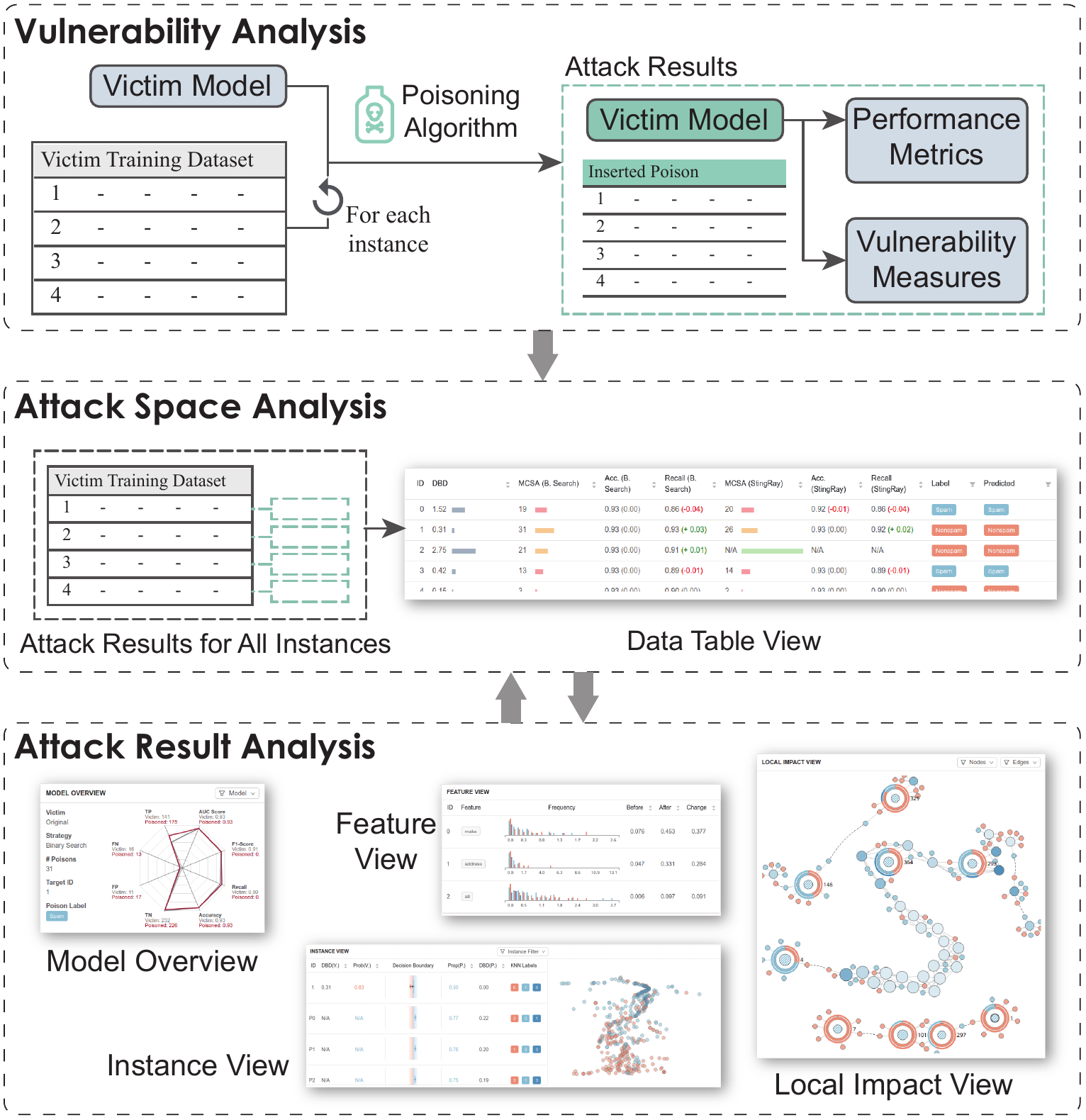}
	\caption{A visual analytics framework for explaining model vulnerabilities to adversarial machine learning attacks. The framework consists of: vulnerability analysis, attack space analysis, and attack result analysis.}
	\label{fig:framework_pipeline}
	\vspace{-5mm}
\end{figure}

Based on the user tasks and design requirements, we have developed a visual analytics framework (Figure~\ref{fig:framework_pipeline}) for identifying vulnerabilities to adversarial machine learning.
The framework supports three main activities: vulnerability analysis, analyzing the attack space, and analyzing attack results. Each activity is supported by a unique set of multiple coordinated views, and the user can freely switch between interfaces and views. All views share the same color mapping in order to establish a consistent visual design. Negative and positive classes are represented by red and blue, respectively, and the dark red and blue colors are used for indicating the labels of poisoning data instances.
All actions in our framework are predicated on the user loading their training data and model. While our framework is designed to be modular to an array of attack algorithms, different performance and vulnerability measures are unique to specific attack algorithms. Thus, for discussion and demonstration, we instantiate our framework on data poisoning attacks.

\subsection{Data-Poisoning Attack Algorithms}
\label{sec:data_poisoning_attack_algorithms}

We focus on the binary classification task described in Figure~\ref{fig:attack_algorithms} (a) where the training data instances are denoted as $\boldsymbol{x} \in \mathcal{X}, \mathcal{X} \subseteq \mathbb{R}^{n \times d}$ with class labels of $y \in \{-1, +1\}$ (we refer to the $-1$ labels as negative and the $+1$ labels as positive). A classification model $\theta$ is trained on the victim training dataset, which creates a victim model. For a target data instance $\boldsymbol{x}_t$ and the corresponding predicted label $y_t = \theta(\boldsymbol{x}_t)$, the attacker's goal is to flip the prediction $y_t$ into the desired class $-y_t$ by inserting $m$ poisoning instances $\mathcal{P} = \{ \boldsymbol{p}_i | \boldsymbol{p}_i \in \mathbb{R}^d, i \in [1, m]\}$. We use $B$ to represent the budget, which limits the upper bound of $m$, i.e., an attacker is only allowed to insert at most $B$ poisoned instances. To maximize the impact of data poisoning on the classifier, the attack algorithms craft poisoned instances in the desired class, $y_{p_i} = -y_t$.

\subsubsection{Attack Strategies}
\begin{figure*}[tbh]
	\centering	
	\includegraphics[width=2.00\columnwidth]{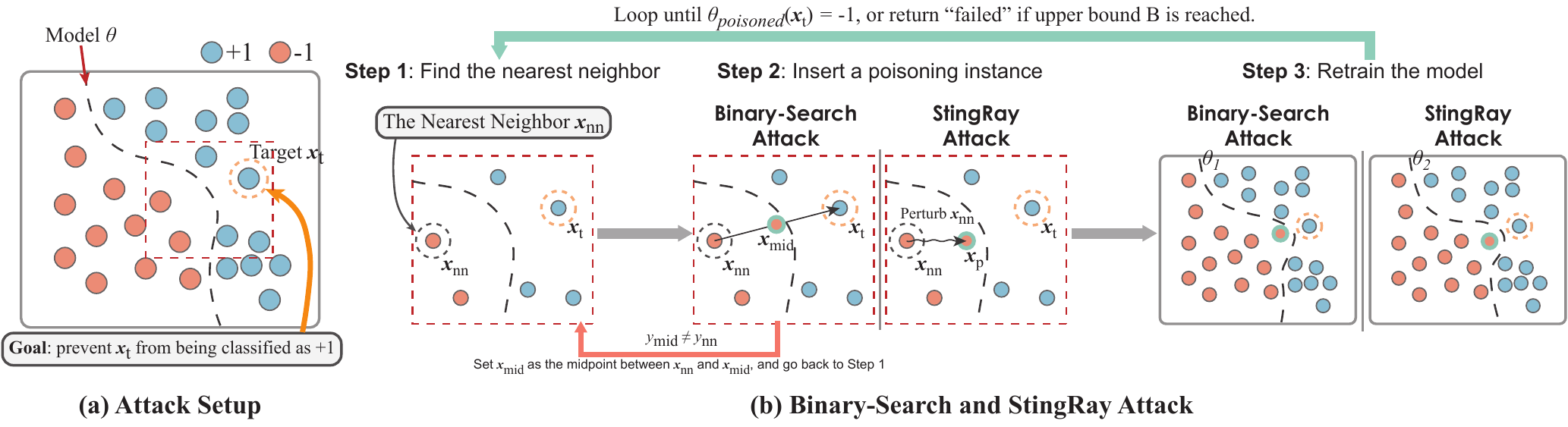}
	\caption{An illustration of data poisoning attacks using the Binary-Search and Stingray algorithms. (a) In a binary classification problem, the goal of a data-poisoning attack is to prevent the target instance, $\boldsymbol{x}_t$, from being classified as its original class. (b) The Binary-Search and StingRay attacks consist of three main steps: 1) select the nearest neighbor to the target instance, 2) find a proper poisoning candidate, and 3) retrain the model with the poisoned training data. The procedure repeats until the predicted class label of $\boldsymbol{x}_t$ is flipped, or the budget is reached.}
	\label{fig:attack_algorithms}
	\vspace{-4mm}
\end{figure*}

Various attack algorithms have been developed to create an optimal set of $\mathcal{P}$ with $|\mathcal{P}| \leq B$. To demonstrate how attacks can be explored in our proposed framework, we implement two different attack algorithms (Binary-Search and StingRay) described in Figure~\ref{fig:attack_algorithms} (b).

\vspace{1.5mm} \noindent \textbf{Binary-Search Attack\footnote{For simplicity, we refer to the Burkard and Lagesse algorithm~\cite{burkard2017analysis} as ``Binary-Search Attack'' even though it is not named by the original authors.}}. The Binary-Search Attack~\cite{burkard2017analysis} assumes that the target instance $\boldsymbol{x}_t$ can be considered as an outlier with respect to the training data in the opposite class $\{\boldsymbol{x}_i | y_i = -y_t\}$. The classification model acts as an outlier detector and separates this target from the opposite class $-y_t$. For crafting poisoning instances in a Binary-Search attack, the goal is to establish connections between the target and the desired class $-y_t$ that mitigate the outlyingness of the target. As illustrated in Figure~\ref{fig:attack_algorithms}, for each iteration, the Binary-Search Attack utilizes the midpoint $\boldsymbol{x}_{mid}$ between $\boldsymbol{x}_t$ and its nearest neighbor $\boldsymbol{x}_{nn}$ in the opposite class, $-y_t$, as a poisoning candidate. If this midpoint is in the desired class, it is considered to be a valid poisoning instance. This instance is appended to the original training dataset, and the model is re-trained ($\theta_1$ in Step 3 - Figure~\ref{fig:attack_algorithms}). In this way, the poisoned instances are iteratively generated, and the classification boundary is  gradually pushed towards the target until the target label is flipped. Sometimes the midpoint may be outside of the desired class. Under this circumstance, a reset of the procedure is required by using the midpoint between $\boldsymbol{x}_{mid}$ and $\boldsymbol{x}_{nn}$ as the new candidate.

\vspace{1.5mm} \noindent \textbf{StingRay Attack}. The StingRay attack~\cite{suciu2018does} inserts new copies of existing data instances by perturbing less-informative features. The StingRay attack shares the same assumptions and pipeline as the Binary-Search attack. The main difference between the attacks is how poisoning instances are generated (Step 2, Figure~\ref{fig:attack_algorithms}). In StingRay, a base instance, $\boldsymbol{x}_{nn}$, near the target, $\boldsymbol{x}_t$, in the desired class is selected, and a copy of the base instance is created as a poisoned candidate. By using some feature importance measures, a subset of features are selected for value perturbation on the poisoned candidate. After randomly perturbing the feature values, the poisoned instance closest to the target is inserted into the training data.

\subsubsection{Attack Results}
Both attacks insert poisoned data instances into the victim training dataset resulting in the \textit{poisoned training dataset}. The model trained on this poisoned dataset is called the \textit{poisoned model}, and we can explore a variety of performance metrics to help explain the results of an attack (e.g., prediction accuracy, recall). For data instance level analysis (D2.2), we derive two metrics that can characterize the impact of data poisoning on the model predictions.

\begin{figure}[tbh]
    \centering	
    \includegraphics[width=1.00\columnwidth]{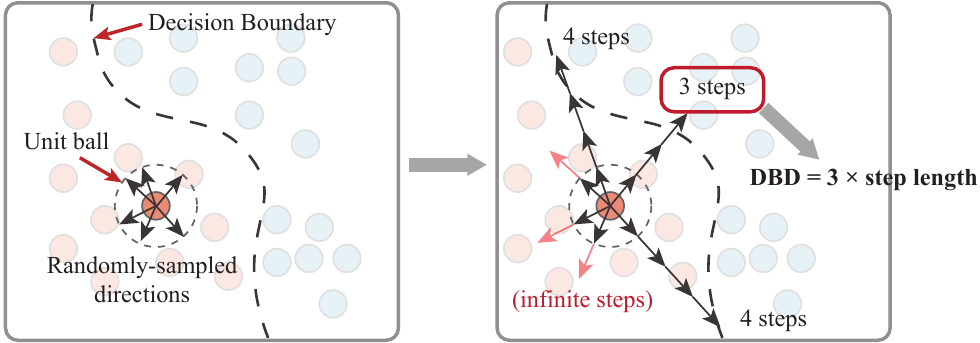}
    \caption{Estimating the decision boundary distance. (Left) Six directional vectors are sampled from the unit ball. (Right) For each direction, the original instance is perturbed one step at a time until it is in the opposite class. In this example, the direction highlighted by the red rectangle is the minimum perturbed step (3 steps) among all the directions.}
    \label{fig:decision_boundary_distance}
	\vspace{-4mm}
\end{figure}

\vspace{1.5mm} \noindent \textbf{Decision Boundary Distance (DBD)~\cite{warren18decision}:} In a classifier, the decision boundary distance is defined as the shortest distance from a data instance to the decision boundary. Under the assumption of outlyingness in the Binary-Search or StingRay attack, DBD is an indication of the difficulty of building connections between a target instance and its opposite class. However, it is difficult (and sometimes infeasible) to derive exact values of DBD from the corresponding classifiers, especially in non-linear models. We employ a sample-based, model-independent method to estimate the DBDs for the training data as illustrated in Figure~\ref{fig:decision_boundary_distance}. First, with a unit ball centered at the data instance, we uniformly sample a set of unit direction vectors from the ball. For each vector, we perturb the original instance along the vector iteratively with a fixed step length, predict the class label with the classifier, and stop if the prediction is flipped. We use the number of perturbation steps as the distance to the decision boundary. We use the product of step length and the minimum steps among all the directions as an estimation of the DBD for each data instance.
    
\vspace{1.5mm} \noindent \textbf{Minimum Cost for a Successful Attack (MCSA):} To help users understand the cost of an attack with respect to the budget, we calculate the minimum number of insertions needed to attack a data instance. For each data instance, the MCSA is the number of poisoning instances that must be inserted for a successful attack under an unlimited budget. The MCSA value is dependent on the attack algorithm.
    
\subsection{Visualizing the Attack Space}
The data table view (Figure~\ref{fig:teaser} (B)) acts as an entry point to the attack process. After loading a model, all the training data instances are listed in the table to provide an initial static check of vulnerabilities (\textbf{T1}, \textbf{D1}). Each row represents a data instance in the training dataset, and columns describe attributes and vulnerability measures which includes the DBD and MCSA for both the Binary-Search and StingRay attack algorithms, as well as the original and the predicted labels. Inspired by Jagielski et al.~\cite{jagielski2018manipulating} and Steinhardt et al.~\cite{steinhardt2017certified}, we use colored bars for MCSA to highlight different vulnerability levels based on the \textit{poisoning rates}, which is defined as the percentage of poison instances in the entire training dataset. Poisoning rates of lower than 5\% are considered to be high risk, since only a small amount of poisoned instances can cause label flipping in these data instances, and poisoning rates of 20\% are likely infeasible (high risk of being caught). 
We define three levels for the poisoning rates: 1) high risk (red) - lower than 5\%; 2) intermediate risk (yellow) - 5\% to 20\%, and; 3) low risk (green) - more than 20\% .

The rows in the table can be sorted by assigning a column as the sorting key. The user can click on one of the checkboxes to browse details on the data ID, class label, and feature values, Figure~\ref{fig:teaser} (B). In addition, the clicking operation will trigger a dialog to choose between the two attack algorithms, and the interface for the corresponding attack result will be opened in a new tab page below. 

\subsection{Visualizing the Attack Results}
After selecting a target instance and an attack algorithm, the user can perform an in-depth analysis of the corresponding attack results. To visualize the results of the attack, we use four views: model overview, instance view, feature view, and $k$NN graph view.

\vspace{1.5mm} \noindent \textbf{Model Overview:} The model overview provides a summary of the poisoned model as well as a comparison between the original (victim) and poisoned model (\textbf{T2}, \textbf{D2.1}). The model overview (Figure~\ref{fig:teaser} (C)) provides a brief summary of the names of the victim and the poisoned models, the ID of the target data instance, and the class of the poisoned instances. A radar chart is used to describe the performance of the two models. The four elements commonly used in confusion matrices (true negative (TN), false negative (FN), true positive (TP), and false positive (FP)) are mapped to the four axes on the left side of the radar chart, and accuracy, recall, F1 and ROC-AUC scores are mapped to the right side. When hovering on the lines, the tooltip shows the detailed values on the axes. The two lines in the radar chart can be disabled or enabled by clicking on the legends.

\vspace{1.5mm} \noindent \textbf{Instance View:}
The instance view illustrates changes in the training datasets and supports the comparative analysis of predictions made by the victim and poisoned models from the perspective of individual data instances (\textbf{T3.1}, \textbf{D2.2}). The instance view is comprised of two sub-views, a projection view and an instance attribute view, which visualize data instances under the ``overview + detail'' scheme.

   \begin{figure}[!ht]
	    \centering	
	    \includegraphics[width=1.00\columnwidth]{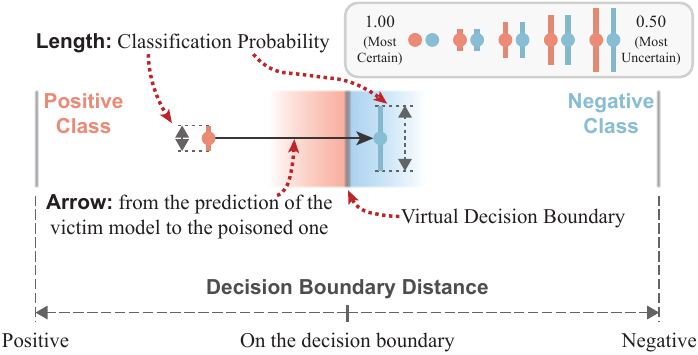}
	    \caption{Design of the virtual decision boundaries in the instance attribute view. The central vertical line acts as the virtual decision boundary. Two circles representing the prediction results of the victim and the poisoned models are placed beside the line. In this example, the data instance far away from the decision boundary was classified as positive with a relatively high probability. However, in the poisoned model, the instance crosses the boundary, causing the label to flip.}
	    \label{fig:instance_attributes_view}
	    \vspace{-5mm}
    \end{figure}

\vspace{1mm} \noindent \textit{Projection View:} The projection view (Figure~\ref{fig:teaser} (D)) provides a global picture of the data distribution, clusters, and relationships between the original and poisoned instances. We apply the t-SNE projection method~\cite{maaten2008visualizing} to the poisoned training dataset. The projection coordinates are then visualized in a scatterplot. We share the colors used in the Model Overview, where red is for label predictions in the negative class and blue for the positive class. To support comparisons between the victim and poisoned model, we apply the corresponding poisoning color to the border of poisoned instances and stripe patterns to the data instances whose class prediction changed after the attack.

\vspace{1mm} \noindent \textit{Instance Attribute View:} The instance attribute view (Figure~\ref{fig:teaser} (E)) uses a table-based layout where each row represents the attributes of an individual data instance including classification probabilities and DBDs from the victim and poisoned model. To conduct a comparison between the attributes of the victim and poisoned models, we embed an illustration of attribute changes into the rows using a virtual decision boundary,  Figure~\ref{fig:instance_attributes_view}. Here, the vertical central line acts as a virtual decision boundary and separates the region into two half panes indicating the negative and positive class regions. Two glyphs, representing the predictions of the victim and the poisoned models, are placed in the corresponding half panes based on the predicted class labels. The horizontal distances from the center dots to the central line are proportional to their DBDs. To show the direction of change, we link an arrow from the victim circle to the poisoned circle. Additionally, the classification probabilities are mapped to the length of the lines in the glyph. A set of options are provided in the top right corner of the view for filtering out irrelevant instances based on their types.

\vspace{1.5mm} \noindent \textbf{Feature View:}
The feature view is designed to reflect the relationship between class features and prediction outputs to help users understand the effects of data poisoning (\textbf{T3.2}, \textbf{D2.3}). In Figure~\ref{fig:teaser} (F), each row in the list represents an individual feature. The feature value distribution is visualized as grouped colored bars that correspond to positive, negative, and poisoning data. To facilitate searching for informative features, the rows can be ranked by a feature importance measure on both the victim and the poisoned models. In our framework, we utilize the feature weights exported from classifiers as the measure, e.g., weight vectors for linear classifiers and Gini importance for tree-based models. In the list, the importance values and their rankings from the two models, as well as the difference, are shown in the last three columns.

\vspace{1.5mm} \noindent \textbf{Local Impact View:}
In order to understand model vulnerabilities, users need to audit the relationship between poisoned instances and targets to gain insights into the impact of an attack (\textbf{T3.1}, \textbf{D2.4}). We have designed a local impact view, Figure~\ref{fig:teaser} (G), to assist users in investigating the neighborhood structures of the critical data instances.

\begin{figure}[ht]
	\centering	
	\includegraphics[width=1.00\columnwidth]{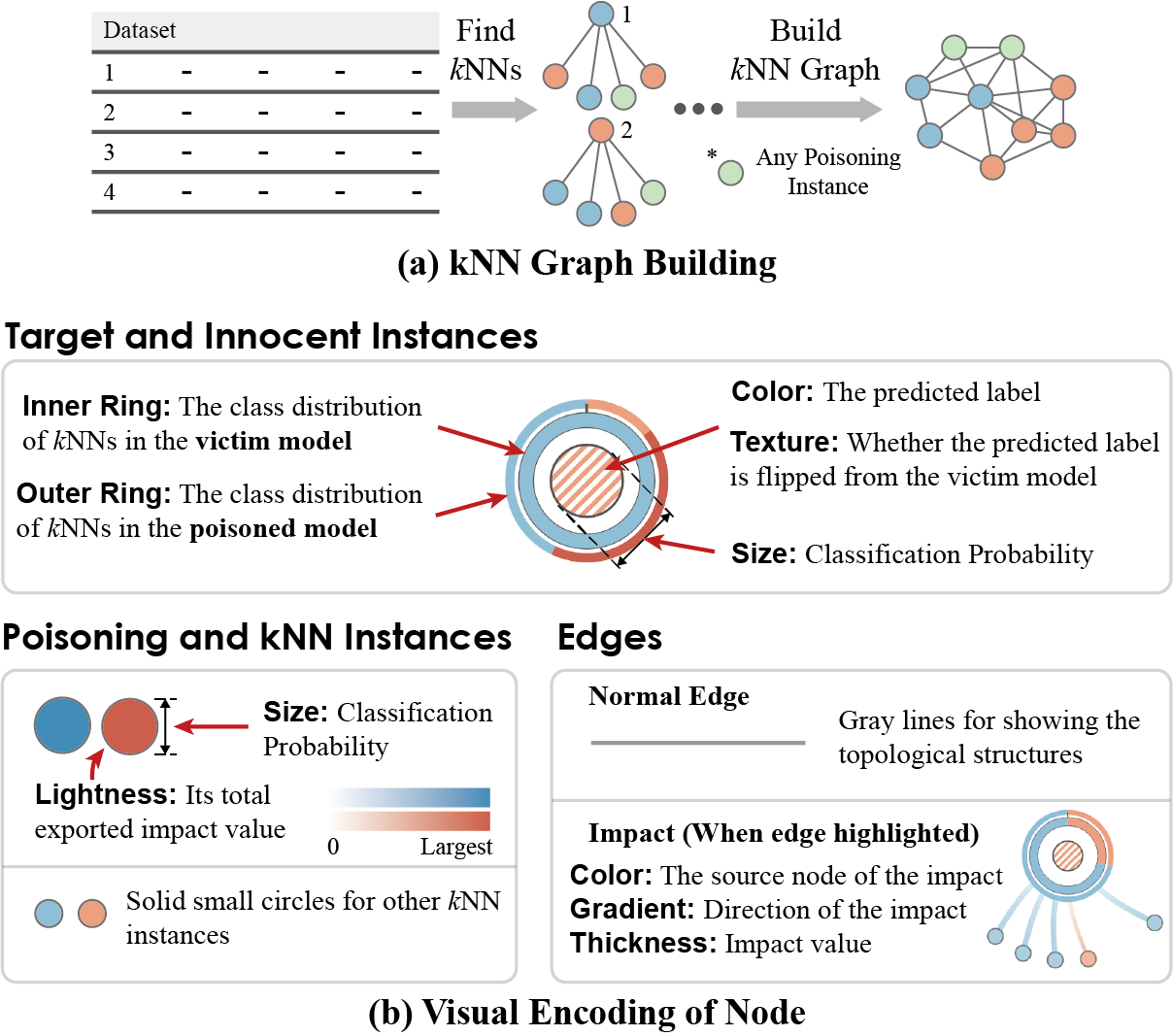}
	\caption{Visual design of the local impact view. (a) The process of building $k$NN graph structures. (b) The visual encodings for nodes and edges.}
	\label{fig:local_glyph_design}
	\vspace{-4mm}
\end{figure}

For characterizing the neighborhood structures of data instances, we utilize the $k$-nearest-neighbor graph ($k$NN graph), Figure~\ref{fig:local_glyph_design} (a), to represent the closeness of neighborhoods, which can reveal the potential impact on the nearby decision boundary. A poisoned instance that is closer to a target may have more impact on the predicted class of the target. Such a representation naturally corresponds to the underlying logic of the attack algorithms, which try to influence the neighborhood structures of target instances. Our view is designed to help the user focus on the most influential instances in an attack. To reduce the analytical burden, we condense the scale of the $k$NN graph to contain only three types of instances as well as their $k$-nearest neighbors:
\begin{enumerate}
    \vspace{-2mm}
    \item The target instance, which is the instance being attacked;
    \vspace{-2mm}
    \item The poisoning instances, and;
    \vspace{-2mm}
    \item The ``innocent'' instances, whose labels are flipped after an attack, which is a side-effect of poisoning.
\end{enumerate}

\vspace{-2mm}
For the target and innocent instances, we extract their $k$NNs before the attack, i.e., the top-$k$ nearest non-poisoned neighbors. This allows the user to compare the two sets of $k$NNs to reveal changes in the local structures after inserting poisoned instances.

The design of the local impact view is based on a node-link diagram of the extracted $k$NN graph where the data instances are represented as nodes. The coordinates of the nodes are computed with the force-directed layout method on the corresponding graph structure. We use three different node glyphs to encode the data instances depending on the instance type (target, poisoned, innocent), Figure~\ref{fig:local_glyph_design} (b).

For the target and innocent instances, we utilize a nested design consisting of three layers: a circle, an inner ring, and an outer ring. The circle is filled with a blue or red color representing the predicted label. A striped texture is applied to the filled color if the label predicted by the poisoned model is different from the victim one, indicating that label flipping has occurred for this data instance. Additionally, the classification probability from the poisoned model is mapped to the radius of the circle. The inner ring uses two colored segments to show the distribution of the two classes in the $k$-nearest non-poisoning neighbors. The outer ring is divided into three segments that correspond to the negative and positive classes and poisoning instances in the $k$NN.

\begin{figure*}[ht]
	\centering	
	\includegraphics[width=\linewidth]{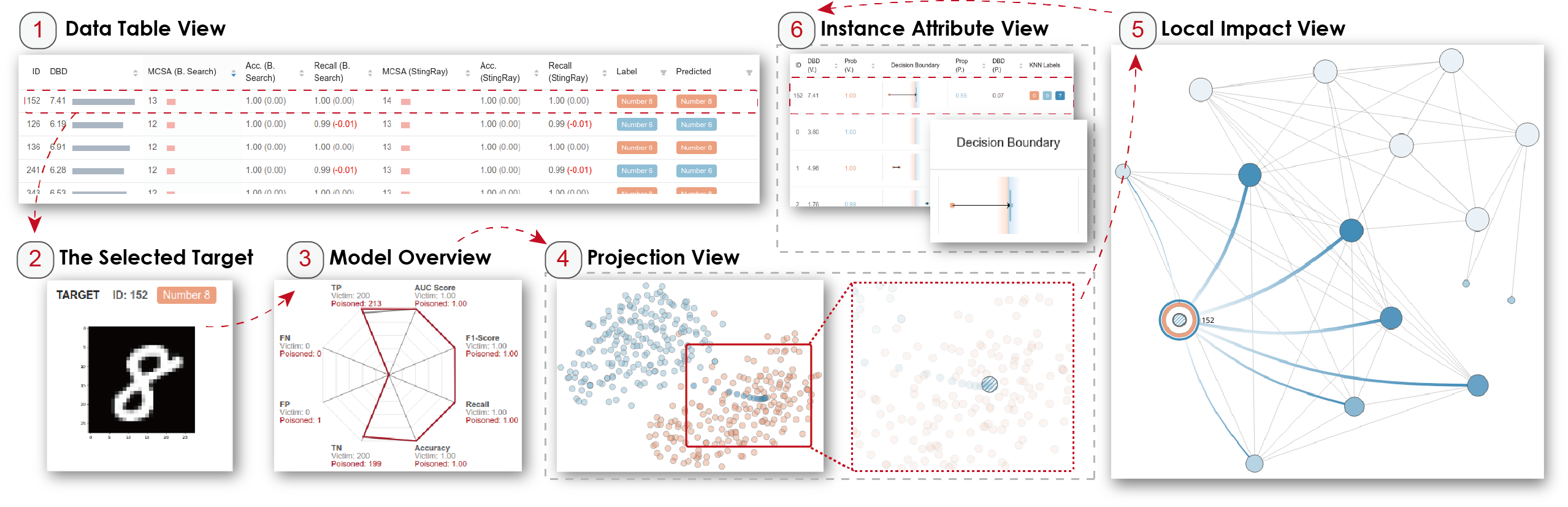}
	\caption{A targeted attack on hand-written digits. (1) In the data table view, we identify the target instance \#152 (2) as a potential vulnerability. (3) In the model overview, we observe no significant change of the prediction performance after an attack on \#152 occurs. (4) In the projection view, the two classes of instances are clearly separated into two clusters. The poisoning instances (dark blue circles) penetrate class Number 8 and reach the neighboring region of instance \#152. (5) The attack can also be explored in the local impact view where poisoning nodes and the target show strong neighboring connections. (6) The detailed prediction results for instance \#152 are further inspected in the instance attribute view.}
	\label{fig:case_study_1}
	\vspace{-4mm}
\end{figure*}

For poisoned instances, we use circles that are filled with the corresponding poisoning color. To depict the total impact on its neighborhoods, we map the sum of the impact values due to poisoned instances to the lightness of the filled color. As in the encoding of the target instances, the radius of the poisoned instance circles represent the classification probability. All other data instances are drawn as small dots colored by their corresponding prediction labels.

The edges in the local impact view correspond to measures of \textit{relative impacts}, which are represented by directed curved edges. Inspired by the classic leave-one-out cross validation method, the relative impact is a quantitative measure of how the existence of a data instance (poisoned or not) influences the prediction result of another instance with respect to the classification probability. Algorithm~\ref{alg:impact} is used to calculate the impact of a neighbor $\boldsymbol{x}_{nn}$ on a data instance $\boldsymbol{x}$. First, we train a new model with the same parameter settings as the poisoned model; however, $\boldsymbol{x}_{nn}$ is excluded. Then, we compute the classification probability of $\boldsymbol{x}$ with this new model. Finally, the relative impact value is calculated as the absolute difference between the new probability and the previous one. To indicate the source of the impact, we color an edge using the same color as the impacting data instance. The color gradient maps to the direction of impact and curve thickness maps to the impact value. Additionally, since the $k$NN graph may not be a fully-connected graph, we employ dashed curves to link the nodes with the minimum distances between two connected components in the $k$NN graph.

\vspace{-4mm}
\begin{algorithm}[htp]
\label{alg:impact}
\SetAlgoLined
\KwData{
    training dataset $\mathcal{X}$; two instances $\boldsymbol{x} \in \mathcal{X}$, $\boldsymbol{x}_{nn} \in \mathcal{X}$; previous classification probability of $\boldsymbol{x}$, $p_{\boldsymbol{x}}$
}

\KwResult{The impact value of $\boldsymbol{x}_{nn}$ on $\boldsymbol{x}$, $I(\boldsymbol{x}_{nn}, \boldsymbol{x})$}

 $\theta \leftarrow \text{Classifier}(\mathcal{X} \setminus \{ \boldsymbol{x}_{nn} \})$
 
 $p'_{\boldsymbol{x}} \leftarrow \text{Probability of } \theta(\boldsymbol{x})$
 
 $I(\boldsymbol{x}_{nn}, \boldsymbol{x}) \leftarrow | p'_{\boldsymbol{x}} - p_{\boldsymbol{x}} |$
 
 \caption{Computing the impact of $\boldsymbol{x}_{nn}$ on $\boldsymbol{x}$}

\end{algorithm}

\vspace{-4mm}

The local impact view supports various interactions on the $k$NN graph. Clicking on a node glyph in the local impact view will highlight the connected edges and nodes and fade out other irrelevant elements. A tooltip will be displayed as well to show the change of neighboring instances before and after the attack. The highlighting effects of data instances are also linked between the projection view and the local impact view. Triggering a highlighting effect in one view will be synchronized in the other one. 

One limitation in the proposed design is the potential for visual clutter once the size of the graph becomes considerably large. In order to provide a clear entry point and support detail-on-demand analysis, we support various filters and alternative representations to the visual elements. By default, the edges are replaced by gray lines, which only indicates the linking relationships between nodes. Users can enable the colored curves mentioned above to examine the impacts with a list of switches, Figure~\ref{fig:teaser} (G.1). Unnecessary types of nodes can also be disabled with the filtering options, Figure~\ref{fig:teaser} (G.2).

\section{Case Study and Expert Interview}

In this section, we present two case studies to demonstrate how our framework can support the analysis of data poisoning attacks from the perspective of models, data instances, features, and local structures. We also summarize feedback from four domain experts.

\subsection{Targeted Attack on Hand-written Digits}
\label{sub:case1}
Digit recognizers are widely-used in real applications including auto-graders, automatic mail sorting, and bank deposits. In such a system, an attacker may wish to introduce reliability issues that can result in mis-delivered mail, or create targeted attacks that cause checks to be mis-read during electronic deposit. For this case study, we employ a toy example in which a model is used to classify hand-written digits. This case study serves as a mechanism for demonstrating system features.

For this classifier, we utilize the MNIST dataset~\cite{lecun1998gradient}, which contains 60,000 images of ten hand-written digits at a 28$\times$28 pixel resolution (784 dimensions in total). We trained a Logistic Regression classifier, implemented in Python Scikit-Learn library~\cite{scikit-learn}, using 200 randomly sampled images from the numbers 6 and 8, respectively. The value of $k$ for extracting $k$NN graphs in the local impact view is set to 7.

\vspace{2mm} \noindent \textbf{Initial Vulnerability Check (T1):} 
After the training dataset and model are loaded into the system, vulnerability measures are automatically calculated based on all possible attacks from the Binary-Search and StingRay Attack, and results are displayed in the data table view (Figure~\ref{fig:case_study_1} (1)). By ranking the two columns of MCSAs for each attack algorithm, the user finds that the red bar colors indicate that many of the data instances are at high risk of a low cost poisoning attack. From the table, the user can also observe that the accuracy and recall values are not highly influenced by an attack, suggesting that a targeted attack on a single instance will not influence prediction performances. To some extent, this may disguise the behavior of a targeted attack by not alerting the model owners with a significant performance reduction.

\vspace{2mm} \noindent \textbf{Visual Analysis of Attack Results (T2, T3):} Next, the user wants to explore a potential worst case attack scenario. Here, they select the instance with the largest MCSA among all the data instances (instance \#152, 3.5\% in poisoning rate) (Figure~\ref{fig:case_study_1} (2)) under the StingRay attack. As illustrated in Figure~\ref{fig:case_study_1} (3), first the user performs a general check of the model performance (\textbf{T2.2}). In the model overview, the two lines on four performance metrics in the radar chart overlap, indicating little to no model performance impact after a poisoning attack. Next, the user explores the distribution of the poisoning instances (\textbf{T2.1, T3.1}). In the projection results, Figure~\ref{fig:case_study_1} (4), the poisoning instances span the border region of two clusters and flip the prediction of the target instance. However, there are no other innocent instances influenced by the poison insertions. The user can further inspect the impact of at attack on instance \#152 by examining the local impact view, Figure~\ref{fig:case_study_1} (5). Here, the user can observe that in a poisoning attack on instance \#152, the neighborhood of \#152 must be heavily poisoned, and these poison insertions establish a connection between the target instance \#152 and two other blue instances, leading to label flipping. In this case, the user can identify that the sparsity of the data distribution in the feature space may be contributing to the vulnerability of instance \#152. Finally, the user explores the detailed prediction result of instance \#152 by navigating to the instance attribute view (Figure~\ref{fig:case_study_1} (6)). Here, the user observes that the label has flipped from Number 8 (red) to Number 6 (blue); however, the poisoning results in a very short DBD and a low classification probability for instance \#152. 

\vspace{2mm} \noindent \textbf{Lessons Learned and Possible Defense:} From the analysis, our domain expert identified several issues in the victim model and dataset. First, even if instance \#152 is successfully poisoned, the instance is fairly near the decision boundary of the poisoned model, which can be identified by the low value of DBD and the low classification probability. If any further data manipulations occur in the poisoned dataset, the prediction of the target instance may flip back, i.e., \#152 is sensitive to future manipulations and the poisoning may be unstable. For the attackers, additional protection methods that mitigate the sensitivity of previous target instances can be adopted by continuously attacking neighboring instances, further pushing the decision boundary away from the target, or improving attacking algorithms to insert duplicated poisons near the target. Our domain expert was also interested in the pattern of a clear connection from the two blue instances to instance \#152 in the local impact view. He noted that it may be due to  data sparsity, where no other instances are along the connection path established by the poisoning instances, resulting in \#152 having a high vulnerability to poisoning insertions. For defenders who want to alleviate the sparsity issue and improve the security of the victim model, possible solutions could be to add more validated labeled samples into the original training dataset and adopt feature extraction or dimension reduction methods to reduce the number of the original features. 

\subsection{Reliability Attack on Spam Filters}
\label{sub:case2}
For spammers, one of their main goals is to maximize the number of spam emails that reach the customers' inbox. Some models, such as the Naive Bayes spam filter, are extremely vulnerable to data poisoning attacks, as known spammers can exploit the fact that the e-mails they send will be treated as ground truth and used as part of classifier training. Since known spammers will have their mail integrated into the modeling process, they can craft poisoned data instances and try to corrupt the reliability of the filter. These specially-crafted emails can mislead the behavior of the updated spam filter once they are selected in the set of new samples. In this case study, we demonstrate how our framework could be used to explore the vulnerabilities of a spam filter.

We utilize the Spambase dataset~\cite{Dua:2019} that contains emails tagged as non-spam and spam collected from daily business and personal communications. All emails are tokenized and transformed into 57-dimensional vectors containing statistical measures of word frequencies and lengths of sentences. For demonstration purposes, we sub-sampled the dataset into 400 emails, keeping the proportion of non-spam and spam emails (non-spam:spam = 1.538:1) in the original dataset, resulting in 243 non-spam instances and 157 spam ones. A Logistic Regression classifier is trained on the sub-sampled dataset. The value of $k$ for the $k$NN graphs is again set to 7.

\vspace{2mm} \noindent \textbf{Initial Vulnerability Check (T1):} Using the Logistic Regression Classifier as our spam-filter model, we can explore vulnerabilities in the training data. For spam filters, the recall score (True-Positives / True-Positives + False-Negatives) is critical as it represents the proportion of detected spam emails in all the ``true'' spams. For a spam filter, a lower recall score indicates that fewer true spam emails are detected by the classifier. We want to understand what instances in our training dataset may be the most exploitable. Here, the user can sort the training data instances by the change in recall score after an attack (Figure~\ref{fig:teaser} (1)). After ranking the two columns of recall in ascending order for each attack algorithm, we found that the Binary-Search attack, when performed on instance \#40, could result in a 0.09 reduction in the recall score at the cost of inserting 51 poisoned instances. 

\vspace{2mm} \noindent \textbf{Visual Analysis of Attack Results (T2, T3):} To further understand what an attack on instance \# 40 may look like, the user can click on the row of instance \#40 and choose ``Binary-Search Attack'' for a detailed attack visualization. In the model overview, Figure~\ref{fig:teaser} (2), we see that the false negative value representing the undetected spams increased from 16 to 30 (nearly doubling the amount of spam e-mails that would have gotten through the filter), while the number of detected spams decreased from 141 to 127. This result indicates that the performance of correctly labeling spam emails in the poisoned model is worse than the victim model (\textbf{T2.2}). 

We can further examine the effects of this attack by doing an instance-level inspection using the projection view (\textbf{T3.1}). As depicted in Figure~\ref{fig:teaser} (3), the two classes of points, as well as the poisoned instances, show a heavy overlap. This indicates that there is an increased possibility of flipping innocent instances coupled with a decrease in prediction performance. In the local impact view (Figure~\ref{fig:teaser} (4)), it can be observed that the poisoning instances are also strongly connected to each other in their nearest neighbor graph (\textbf{T2.1}). Additionally, there are five poisoning instances with a darker color than the others. As the lightness of poisoning nodes reflects their output relative impact, these five neighbors of the target instance contributes more to the prediction results than other poisons. For target instance \#40, the outer ring consists only of the poisoning color, indicating that it must be completely surrounded by poisoning instances in order for the attack to be successful. Additionally, in a successful attack, there would be more than 20 innocent instances whose label are flipped from spam to non-spam, which is the main cause of the decreased recall value. After examining the details of these instances in Figure~\ref{fig:teaser} (5), we found that most of their DBDs in the victim model are relatively small, i.e., they are close to the previous decision boundary. As such, their prediction can be influenced by even a small perturbation of the decision boundary. Finally, we conducted a feature-level analysis by browsing the feature view (\textbf{T3.2}, Figure~\ref{fig:teaser} (6)). We find that for distributions of poisoning instances along each feature, the variances are quite large on some words including ``will'' and ``email''. This suggests that there are large gaps between the non-spam emails and instance \#40 on these words in terms of word frequencies, which could be exploited by attackers when designing the contents of the poisoned emails.

\vspace{1mm} \noindent \textbf{Lessons Learned and Possible Defense:} From our analysis, our domain expert was able to identify several key issues. First, from the distribution of impact values and classification probabilities among the poisoning instances, an interesting finding was that the poisoning instances close to the target are more uncertain (i.e., of low classification probability values) and essential to flipping its label. Our domain expert mentioned that further optimization may be performed by removing poisoning instances far away from the target because their impact and classification uncertainty could be too low to influence the model training.
Second, even though an attack on instance \#40 has the maximum influence on the recall value, there is a large (but not unfathomable) cost associated with inserting 51 poisoning instances (poisoning rate = 12.75\%). Given the large attack cost, our domain expert was interested in exploring alternative attacks with similar impacts and lower costs, such as instance \#80 (Figure~\ref{fig:teaser} (7)). A poisoning attack on \#80 can result in a reduction of 0.07 on the recall at almost half the cost of \#40 (29 insertions, poisoning rate = 7.25\%). The key takeaway that our analyst had was that there are multiple viable attack vectors that could greatly impact the reliability of the spam filter. Given that there are several critical vulnerable targets, the attackers could perform continuous low-cost manipulations to reduce the reliability of the spam filter. This sort of approach is typically referred to as a ``boiling-frog attack~\cite{Adv2018book}''. Here, our domain expert noted that the training-sample selection process may need to be monitored.

\subsection{Expert Interview}
To further assess our framework, we conducted a group interview with our collaborator (E0) and three additional domain experts in adversarial machine learning (denoted as E1, E2, and E3). For the interview, we first introduced the background and goals of our visual analytics framework, followed by an illustration of the functions supported by each view. Then, we presented a tutorial of the analytical flow with the two case studies described in Section~\ref{sub:case1} and ~\ref{sub:case2}. Finally, the experts were allowed to freely explore the two datasets (MNIST and Spambase) in our system. The interview lasted approximately 1.5 hours.

At the end of the interview session, we collected free-form responses to the following questions:

\begin{enumerate}
\item Does the system fulfill the analytical tasks proposed in our work?
\item Does our analytical pipeline match your daily workflow? 
\item What are the differences between our visual analytics system and conventional machine learning workflows? 
\item Is the core information well-represented in the views? 
\item Are there any views that are confusing, or that you feel could have a better design?
\item What results can you find with our system that would be difficult to discover with non-visualization tools? 
\end{enumerate}

\vspace{2mm}\noindent \textbf{Framework:} The overall workflow of our framework received positive feedback with the experts noting that the system was practical and understandable. E3 appreciated the two-stage (attack space analysis and attack result analysis) design in the interface, and he conducted a combination of ``general checks + detailed analysis''. E2 noted that \textit{``the stage of attack space analysis gives our domain users a clear sense about the risk of individual samples, so we can start thinking about further actions to make the original learning models more robust and secure,''}. E1 mentioned that the framework could be easily adapted into their daily workflow and improve the efficiency of diagnosing new poisoning attack algorithms. E1 also suggested that it will be more flexible if we can support hot-swapping of attack algorithms to facilitate the diagnosis process.

\vspace{2mm}\noindent \textbf{Visualization:} All the experts agreed that the combination of different visualization views can benefit multi-faceted analysis and provide many aspects for scrutinizing the influence of poisoning attacks. E2 was impressed by the instance attribute view and felt that the glyphs were more intuitive than looking at data tables since the changes of distances to the decision boundary can be directly perceived. E3 mentioned that the local impact view provides essential information on how the neighboring structures are being influenced. The two-ring design of the target and innocent instances provides a clear comparison of two groups of nearest neighbors before and after an attack. E3 further added that the node-link diagram and the visual encoding of impacts are effective for tracing the cause of label flipping and the valuable poisoning instances. \textit{``With the depiction of impacts, maybe we could find how to optimize our attack algorithms by further reducing the number of insertions, since some of the low-impact poisoning instances may be removed or aggregated.''}

\vspace{2mm}\noindent \textbf{Limitations:} One issue found by our collaborator, E0, was the training time that was necessary for using our framework. E0 commented that during the first hour of the interview, we were often required to repeat the visual encoding and functions in the views. However, once the domain experts became familiar with the system after free exploration for some time, they found that the design is useful for gaining insights from attacks. We acknowledge that there could be a long learning curve for domain experts who are novice users in comprehensive visual analytics systems.

\section{Discussion and Conclusions}
In this work, we propose a visual analytics framework for exploring vulnerabilities and adversarial attacks to machine learning models. By focusing on targeted data poisoning attacks, our framework enables users to examine potential weak points in the training dataset and explore the impacts of poisoning attacks on model performance. Task and design requirements for supporting the analysis of adversarial machine learning attacks were identified through collaboration with domain experts. System usability was assessed by multiple domain experts and case studies were developed by our collaborating domain scientist. Target users of our framework are data scientists who utilize machine learning models in mission-critical domains. In contrast to traditional reactive defense strategies that respond when attacks are detected, our framework serves as a mechanism for iterative proactive defense. The users can simulate poisoning operations and explore attack vectors that have never been seen in the historical records. This can enable domain scientists to design more reliable machine learning models and data processing pipelines. An implementation of our framework is provided in Github\footnote{\url{https://github.com/VADERASU/visual-analytics-adversarial-attacks}}.

\vspace{2mm} \noindent \textbf{Target Users:} The target users of our framework are data scientists and security experts who wish to explore model vulnerabilities. Data scientists can use the proposed framework to perform extensive checks on their model training processes in order to enrich the quality of training datasets. Similarly, security experts can benefit from using our framework by actively adopting new attack strategies for the purpose of penetration testing following the paradigm of ``security-by-design'' in proactive defense~\cite{biggio2018wild}.

\vspace{2mm} \noindent \textbf{Limitations:} One major concern in our design is scalability. We have identified issues with both the attack algorithms and visual design.

\vspace{1mm} \noindent \textit{Attack Algorithms:} The computational efficiency of an attack algorithm has a significant influence on the cost of pre-computing vulnerability measures. In order to explore vulnerabilities in data poisoning, every data instance must undergo an attack. In the two case studies, it takes about 15 minutes to compute the MCSA values for the 400 training data instances. In large-scale datasets, this may make the pre-computation infeasible. Sampling methods could be used to reduce the analysis space, and weighted sampling can be adopted to increase the number of samples in the potentially vulnerable regions in the feature space. An upper limit on attack costs could also be used so that a poisoning attack would simply be marked as ``failure" if the upper bound is reached. Furthermore, progressive visual analysis~\cite{stolper2014progressive,angelini2018review} can be employed in the sampling process, allowing users to conduct a coarse-grained analysis of the samples and then increase sample rates on targeted regions.

\vspace{1mm} \noindent \textit{Visual Design:} In our visual design, circles may overlap when the number of training data instances is over one thousand. To mitigate visual clutter in the projection result, we have employed semantic zooming in the projection view to support interactive exploration in multiple levels. In the future, various abstraction techniques for scatterplots such as Splatterplots~\cite{Mayorga2013}, down-sampling~\cite{Chen2014Bluenoise}, and glyph-based summarization~\cite{Liao2018} can be integrated to reduce the number of points displayed in the canvas. Interactive filtering can also be adopted to remove the points in less important regions, e.g., far away from the target instance. A similar issue also exists in the local impact view where the current implementation supports up to 100 nodes shown as graph structures. The readability of large graph visualizations is still an open topic in the community. One way to scale our design would be to build hierarchical aggregation structures on the nodes by clustering the corresponding instances with specific criteria~\cite{freire2010manynets,2012-graphprism,Yoghourdjian2018}. 

\vspace{2mm} \noindent \textbf{Future Work:} In this work, we use the data poisoning attack as the main scenario to guide the visual design and interactions in the visual analytics framework. Based on the successful application in poisoning attacks, we plan to adapt our framework for other typical adversarial settings and attack strategies, such as label-flipping attacks~\cite{steinhardt2017certified} where the labels of training data instances can be manipulated, and evasion attacks~\cite{dalvi2004adversarial,10.1007/978-3-642-40994-3_25,43405} that focus on the testing stage. Another issue of our work is that the framework currently does not support the integration of known defense strategies. In practice, attack and defense strategies often co-exist and must be simultaneously considered in assessing vulnerabilities. Future iterations of this framework will incorporate defense methods as a post-processing stage to evaluate the vulnerability and effectiveness of attacks under countermeasures. In addition, since currently our work only considers classification models for general-purpose tasks, another extension would be to specialize our framework to support domain-specific analyses, such as image recognition, biological analysis, and network intrusion detection.

\acknowledgments{
This work was supported by the U.S.Department of Homeland Security under Grant Award 2017-ST-061-QA0001. The views and conclusions contained in this document are those of the authors and should not be interpreted as necessarily representing the official policies, either expressed or implied, of the U.S. Department of Homeland Security.
}
\bibliographystyle{abbrv-doi}
\bibliography{template}
\end{document}